# Scale-free vortex cascade emerging from random forcing in a strongly coupled system

K. Rypdal*[1], B. Kozelov[1,4], S. Ratynskaia[2], B. Klumov[3], & C. Knapek[3]

[1] *Department of Physics and Technology, University of Tromsø, 9037 Tromsø, Norway*

[2] *Alfven Laboratory, Royal Institute of Technology, Stockholm, Sweden*

[3] *Max-Planck Institute of Extraterrestrial Physics, Garching, Germany*

[4] *Polar Geophysical Institute, Kola Science Centre of Russian Academy of Sciences, Apatity, Russia*

The notions of self-organised criticality (SOC)[1] and turbulence[2] are traditionally considered to be applicable to disjoint classes of phenomena. Nevertheless, scale-free burst statistics is a feature shared by turbulent[3] as well as self-organised critical dynamics[4]. It has also been suggested that another shared feature is universal non-gaussian probability density functions (PDFs) of global fluctuations[5,6]. Here, we elucidate the unifying aspects through analysis of data from a laboratory dusty plasma monolayer[7]. We compare analysis of experimental data with simulations of a two-dimensional (2D) many-body system[8], of 2D fluid turbulence[9], and a 2D SOC model[10], all subject to random forcing at small scales. The scale-free vortex cascade is apparent from structure functions as well as spatio-temporal avalanche analysis[4], the latter giving similar results for the experimental and all model systems studied. The experiment exhibits global fluctuation statistics consistent with the universal PDF described in Refs. 5,6 , but the model systems yield this result only in a restricted range of forcing conditions.

In the experimental setup 600 micron-size charged dust grains immersed in a weakly ionized plasma levitate above an electrode in a radio-frequency plasma discharge and form states ranging from hexagonal 2D crystals to 2D liquid or gaseous states. The most interesting states, however, are those that are semi-crystalline, but show strong grain transport and even vortical viscoelastic flows and turbulence due to development of defects in the crystal structure[7]. The particles are illuminated by a laser beam, their positions are recorded by a CCD camera, and their trajectories can be tracked to yield the full space-time history of every particle in the horizontal plane. The experimental setup and methods are described in Ref. 7.

A natural unit of collective motion is a hexagonal vortex as seen in Fig. 1a, but topological restrictions do not allow a system of hexagonal cells, since such a system requires counter-streaming flows along some edges of neighbouring vortices. This velocity shear creates stress leading to break-up and merging of vortices. The velocity field of the resulting turbulent dynamics can be visualised (Fig. 1a) and animated by plotting the particle positions in time in grey-scale plots with decaying light intensity backwards in time. Vortical structures lasting shorter than the chosen decay time will not be visible in this animation. However, vortex dynamics on all scales down to the sampling time and the mean inter-particle distance can be visualised by creating a mean velocity field $\mathbf{v}(\mathbf{x},t)$ by spline interpolation of the particle velocities in the neighborhood of any point $\vec{x}$ in the plane. Scalar field of the mean particle speed $U(\mathbf{x},t) = |\mathbf{v}(\mathbf{x},t)|$, kinetic energy $K = mU^2/2$, or vorticity $\omega = |\nabla \times \mathbf{v}|$ can also be constructed. In Fig. 1c the field $U(\mathbf{x},t)$ is plotted in a grey-scale plot. The corresponding animations appear as a boiling soup where the smallest blobs retain their identity for times shorter than the decay time defining the velocity field plots in Fig . 1 a,b. Similar plots and animations can be made for vorticity.

In the many-body simulation one follows the trajectories of charged point masses subject to self-consistent, mutually repulsive electrostatic forces, in addition to the confining force from a parabolic confining potential. The effect of collisions with the particles of the neutral background gas has been modelled by a stochastic force with zero mean and a friction force proportional to the velocity of the point mass.

Data from the experiment has been used to model these forces as realistically as possible, but there are limitations in this respect. The main difference between the experimental and simulated dynamics is that the ratio between "thermal" particle velocities and large-scale fluid velocities are higher in the simulation than in the experiment. However, the most interesting features of the scale-free vortex cascade in the experimental flow are reproduced by the simulation, as will be shown below.

Spatial characteristics of the flow have some commonalities with 2D Navier-Stokes turbulence. A standard method of characterizing scaling in turbulence is to compute structure functions $S_m(d) = \langle |\mathbf{v}(\mathbf{r}+\mathbf{d}) - \mathbf{v}(\mathbf{r})|^m \rangle \sim d^{\xi(m)}$. In Fig. 2 a,b we show in a double-logarithmic plot structure functions of increasing order of the particle flow-field in the core of the dust-particle cluster (experiment and simulation), and how the slopes $z(m) = \zeta(m)/m$ of these structure functions depend on the order $m$. If $z$ is independent of $m$ we have self-similar scaling. For the experiment the results indicate two scaling regimes: for spatial scales less than about 10 mean inter-particle distances the crystalline order gives rise to oscillations in the structure function and the similarity exponent calculated from a fitted straight line is $z \approx 0.1$. For longer scales we have a different scaling regime with $z \approx 0.25$, which is due to an azimuthal "zonal flow" component. For comparison $z = 1/3$ from Kolmogorov's K41 theory for isotropic turbulence[2]. The many-body simulations do not exhibit these zonal flows and, as shown in Fig. 2b, do not exhibit the scaling regime with $z \approx 0.25$ for large $d$, but has scaling similar to the experiment for $d$ less than 10 inter-particle distances.

Dust grains are partially trapped in their potential wells in the quasi-crystal, and their motion over longer distances than the mean inter-grain separation occurs through an avalanche of defects in the crystal structure. Avalanches (or bursts) can be identified as a connected cluster of particles in 3D space-time whose speed exceeds a prescribed threshold. This definition associates avalanching particles with those in the edge region of vortices. In space-time such clusters may be cylinder-shaped. However, since vortices split and merge, avalanches generally constitute tree-like structures in space-time (Fig. 3a). Let the avalanches in a given observation series (or a simulation) be enumerated by the index $i = 1,2,...,N$ according to their time of birth $t_i$, where $t_1 \leq t_2 \leq ... \leq t_N$, and assume that these avalanches die at times $T_i$, $i = 1,2,...,N$. The instantaneous avalanche area $a_i(t)$ is the number of avalanching particles at time $t$ (the intersection of the avalanche cluster with the $t$-plane) and the avalanche size $A_i = \sum_{t=t_i}^{T_i} a_i(t)$ is the total number of particles in such a space-time cluster. The avalanche duration $\tau_i = T_i - t_i$ is the cluster extent in the time direction.

For a system in a self-organised critical state the PDFs of avalanche size, duration and area take the form of power laws. Let us assume that the probability that an avalanche lives longer than $\tau$ scales as $F_\tau(\tau) \sim \tau^{-\delta}$ and that the avalanche area disperses with time as $a \sim t^h$. The size of an avalanche with duration $\tau$ will then scale as $A(\tau) = \int_1^\tau t^h dt \sim \tau^{h+1}$ and $dA \sim \tau^h d\tau$. Assuming that the PDF of avalanche sizes has the form $P_A(A) \sim A^{-\nu}$ the relation $P_A(A) dA = (dF_\tau / d\tau) d\tau$ yields $\tau^{-\nu(h+1)+h} \sim \tau^{-\delta-1}$, and hence we get the relation between the characteristic exponents, $\nu = (h + \delta + 1)/(h + 1)$.

In Fig. 3b we plot the mean area $\langle a(t) \rangle \sim t^h$ of avalanches living longer than the time $t$, and find $h$ for the avalanche area stochastic process of the dust monolayer experiment and simulation. In Fig. 3c the plots of the survival probability $F(\tau)$ yield $\delta$, and in Fig. 3d the plots of the avalanche size distribution $P_A(A)$ yield $\nu$. These exponents are consistent with the relation between exponents given above, for experiment as well as simulations.

We have tested these results against simulations of 2D Navier-Stokes turbulence where energy is injected by random perturbations of the velocity field at small scales (see Methods section), and simulations of the Zhang sandpile model[13]. For these simulations avalanche analysis can be done in the same manner as for the dusty monolayer experiment and simulation. The characteristic exponents for these four different systems derived from the avalanche analysis are summarised in Table 1.

One of the great mysteries of the unity of turbulent and avalanching dynamics is the mechanism behind the claimed universal PDFs of global quantities. Analysis we have performed on data from the dust experiments show that the PDF of global fluctuations in kinetic energy fits the so-called Bramwell-Holdsworth-Pinton (BHP) distribution[5,6] quite well. The same turns out to be true for the kinetic energy fluctuations of the 2D fluid simulation and for the toppling activity fluctuations in the Zhang-model, for some simulation conditions. These results are shown in Figure 4, and may be taken as support of the idea of a universal PDF of global fluctuations in turbulent and critical systems. However, simulations we have made on weakly and strongly driven sandpile models show that this form is only well described by the BHP form for a moderately strong drive. For a weak drive the PDF is a stretched exponential, and for strong drive it approaches a more symmetric form. We also find that the kinetic energy PDF for the fluid simulations depends on the details of the forcing. These rather severe limitations of the universality of the BHP distribution need to be considered in future theoretical efforts to explain its appearance.

2D turbulence is very different from its 3D counterpart in that it exhibits an inverse energy cascade to larger scales, giving rise to merging of vortices and generation of large-scale flows. This is because the phenomenon of vortex stretching is impossible in 2D. Emergence of system-size vortices has its counterpart in system-size avalanches in SOC models. The many-body simulations of the dusty monolayer demonstrate that a scale-free distribution of vortices may result from stochastic, independent forcing of each individual dust grain. This is the essence of self-organised criticality: a scale-free hierarchy of dynamical entities emerging in an open, non-equilibrium system subject to a random drive on the smallest possible spatial scales. These apparent commonalities between 2D turbulence and sandpile avalanching, and their physical realisation in the dust monolayer, call for more general approaches to turbulence which go beyond the Navier-Stokes equation and invoke concepts from the theory of critical phenomena in non-equilibrium systems[11].

## REFERENCES

1. Jensen, H. J., Self-organized criticality. Emergent complex behavior in physical and biological systems (Cambridge University Press 1998).

2. Frisch, U., Turbulence: the legacy of A. N. Kolmogorov, Cambridge University Press, New York (1995).

3. Boffetta, G, Carbone, V., Giuliani, P., Veltri, P., & Vulpiani, A., Power-laws in solar flares: self-organized criticality or turbulence? *Phys. Rev. Lett.* **83**, 4662 - 4665 (1999).

4. Paczuski, M., Boettcher, S., & Baiesi, M., Interoccurrence times in the Bak-Tang-Wiesenfeldt sandpile model: a comparison with the observed statistics of solar flares. *Phys. Rev. Lett.* **95**, 181102-1-4 (2005).

5. Bramwell, S. T., Holdsworth, P. C. W., & Pinton, J.-F., Universality of rare fluctuations in turbulence and critical phenomena, *Nature* **396**, 552-554 (1998).

6. Bramwell, S. T., Christensen K., Fortin, J.-Y., Holdsworth, P. C. W., Jensen, H. J., Lise, S., Lopez, J. M., Nicodemi, M., Pinton, J.-F., & Sellitto, M., Universal fluctuations in correlated systems, *Phys. Rev. Lett.* **84**, 3744-3747 (2000).

7. Ratynskaia, S., Rypdal, K., Knapek, C. et al., Superdiffusion and viscoelastic vortex flows in a two-dimensional complex plasma, Phys. Rev. Lett. **96**, 105010 (2006).

8. Klumov B.A., Rubin-Zuzic M., Morfill G.E., Crystallization waves in a dusty plasma, JETP LETTERS 84 (10), p542-546, (2007).

9. Babianao, A., Dubrulle, B., & Frick, P., Scaling properties of numerical two-dimensional turbulence, Phys. Rev. E., **52,** 3719 (1995).

10. Zhang, H. Y., Scaling theory of self-organized criticality, Phys. Rev. Lett. **63**, 470 (1988).

11. Sornette, D., Critical Phenomena in Natural Sciences (Springer 2000).

12. Hoshen, J., & Kopelman, R., Percolation and cluster distribution: Cluster multiple labeling technique and critical concentration algorithm, Phys. Rev. B, V.14, P.3438–3445 (1976).

13. Stam, J., Stable Fluids. In Proceedings of SIGGRAPH (1999). http://www.dgp.toronto.edu/people/stam/reality/Research/pdf/ns.pdf

14. Hnat, B., Chapman, S.C., Rowlands, G., Watkins, N.W., & Freeman, M.P., Scaling in long term data sets of geomagnetic indices and solar wind ε as seen by WIND spacecraft, Geophys. Res. Lett., 30, 2174-2177 (2003).

15. Kovács, P., Carbone V., & Vörös Z., Wavelet-based filtering of intermittent events from geomagnetic time series, Planet.Space Sci., 49, 1219-1231 (2001).

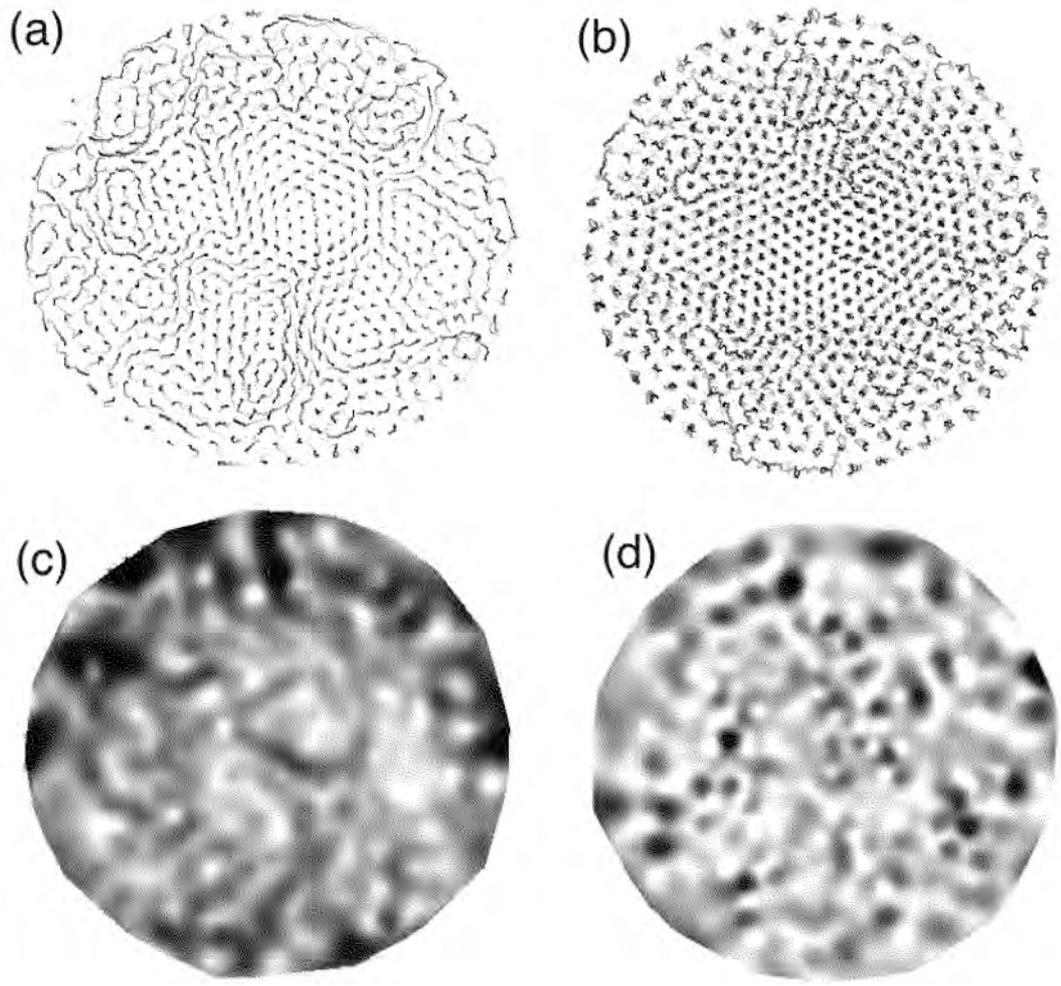

**Figure 1 | Field of dust grain velocity and speed in dust monolayer experiment (left) and in many-body simulation (right).**

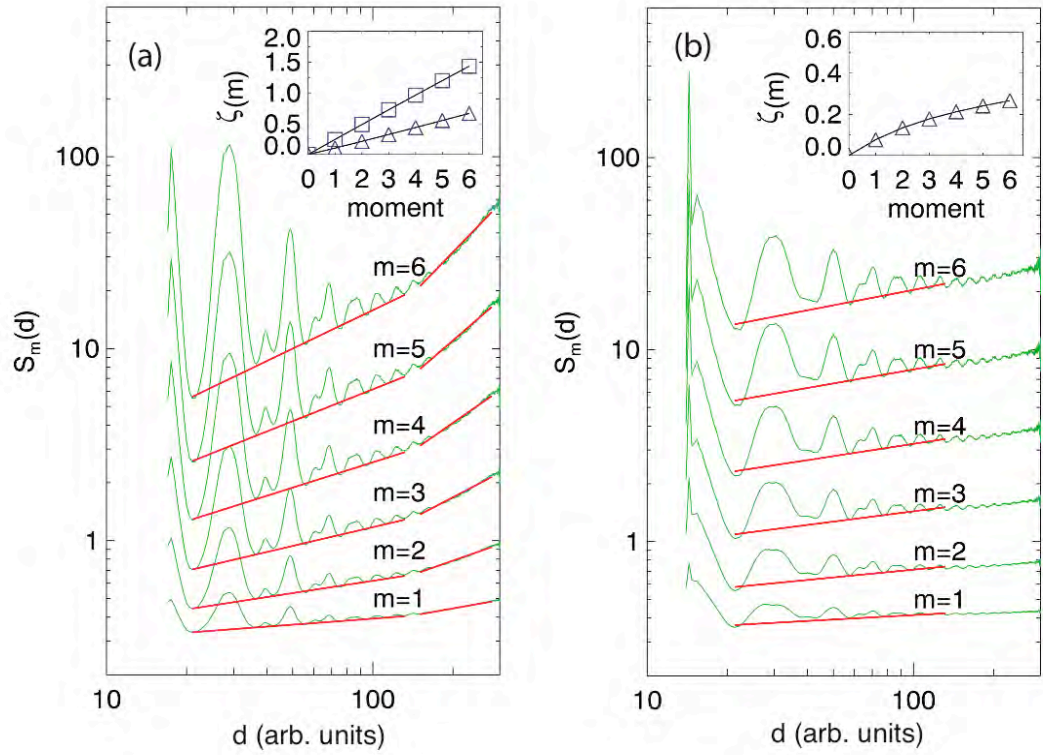

**Figure 2 | Structure functions and estimation of self-similarity exponent for dust monolayer experiment (left) and for many-body simulation (right).**

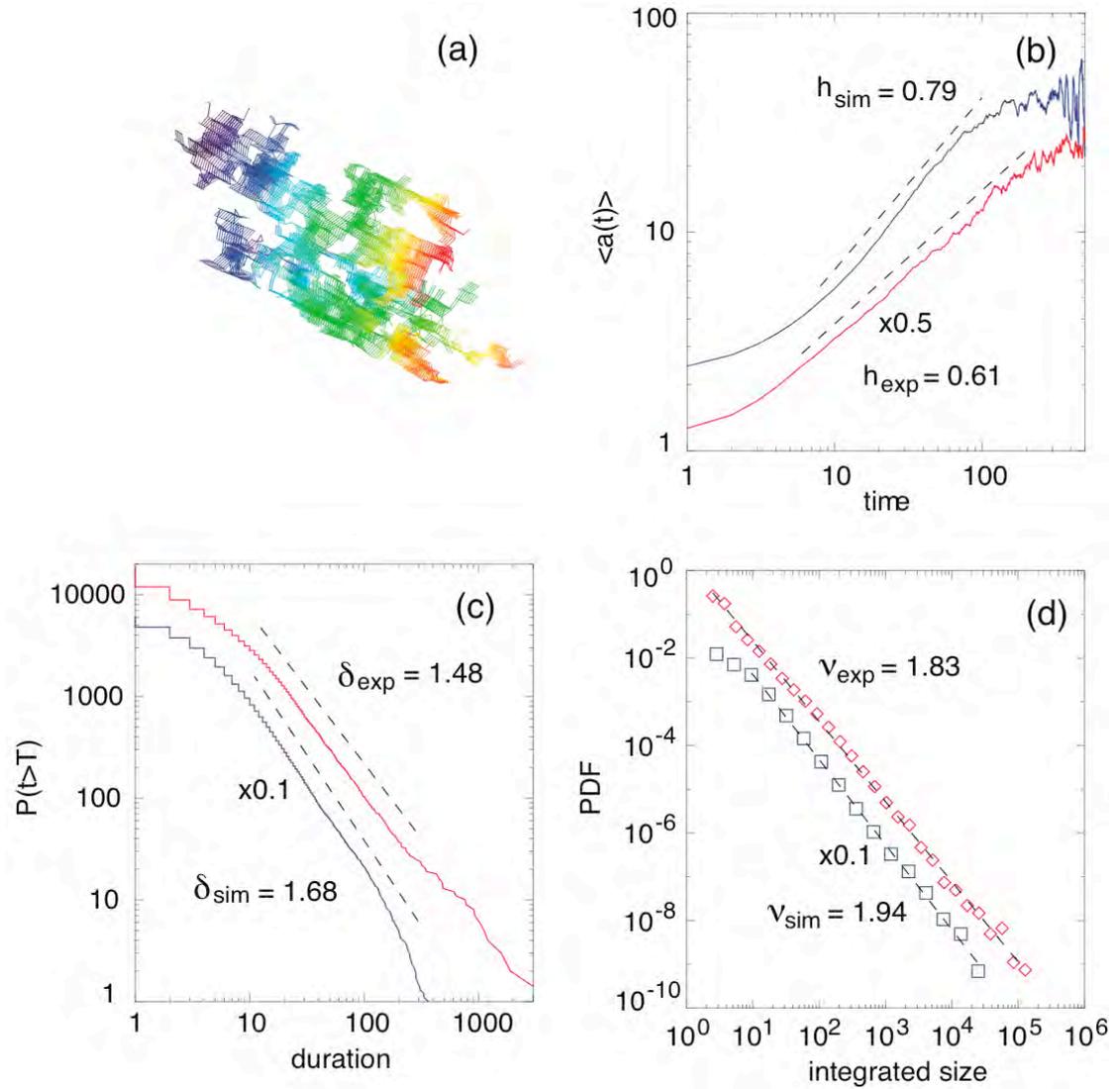

**Figure 3 | Avalanche analysis of dust monolayer experiment and of many-body simulation.** a, The 3D space-time structure of a large avalanche in the dust experiment. Time runs along the axis of the elongated structure, blue representing the start – and red the end of the avalanche. In figures b, c, and d, the red curves refer to experiment and the blue curves to simulation.

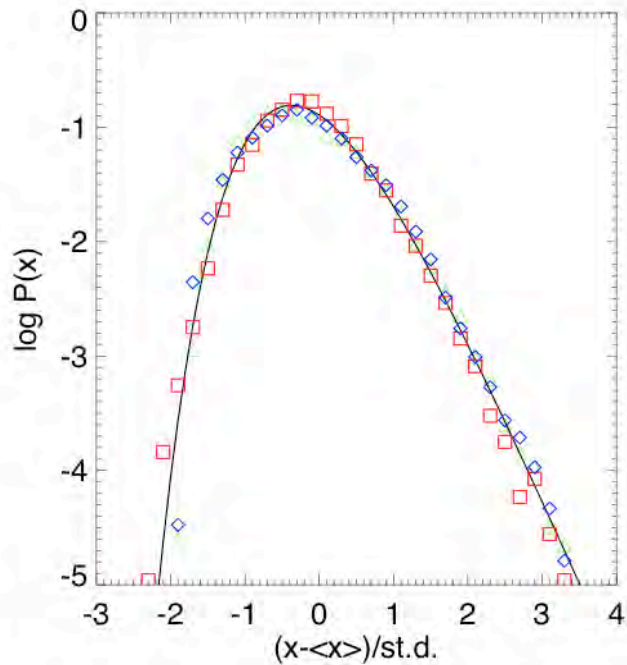

**Figure 4 | PDFs of global fluctuations.** PDFs of fluctuations in total kinetic energy for dust experiment (triangles), for 2D fluid turbulence simulation (squares), and for total toppling activity in the Zhang sand pile model with moderately strong drive (diamonds). The full curve is the BHP distribution.

**Table 1 | Critical exponents for dust experiment and simulation, 2D fluid turbulence simulation, and for total toppling activity in the Zhang sandpile model with moderately strong drive.**

|  | $\delta$ | $\nu$ | h | $\nu^*$ |
|---|---|---|---|---|
| Dusty experiment | 1.48±0.02 | 1.83±0.01 | 0.61±0.02 | 1.92 |
| Dusty simulation | 1.68±0.02 | 1.94±0.01 | 0.79±0.02 | 1.94 |
| 2D fluid simul. | 1.34±0.01 | 1.78±0.03 | 0.77±0.02 | 1.76 |
| 2D Zhang model | 1.26±0.01 | 1.45±0.01 | 1.55±0.01 | 1.49 |

*estimated from the relation $\nu=(h+\delta+1)/(h+1)$